\documentclass[10pt]{article}

\usepackage[T1]{fontenc}
\usepackage[utf8]{inputenc}
\usepackage[english]{babel}
\usepackage{lmodern}
\usepackage{amsmath,amssymb}
\usepackage{graphicx}
\usepackage{geometry}
\usepackage{microtype}
\usepackage{xurl}
\usepackage[hidelinks]{hyperref}
\usepackage[super,sort&compress]{natbib}
\usepackage{caption}
\usepackage{enumitem}
\usepackage{titlesec}
\usepackage{placeins}
\usepackage{float}

\graphicspath{{Compression_extension_asymmetry_manuscript_media/}}
\newcommand{\manuscriptfigure}[3][\linewidth]{%
  \IfFileExists{Compression_extension_asymmetry_manuscript_media/#3}{%
    \includegraphics[width=#1,height=#2,keepaspectratio]{#3}%
  }{%
    \fbox{\parbox[c][0.20\textheight][c]{0.94\linewidth}{\centering
      External figure asset unavailable in this isolated preview.}}%
  }%
}
\titleformat{\section}{\large\bfseries}{\thesection}{0.55em}{}
\titleformat{\subsection}{\normalsize\bfseries}{\thesubsection}{0.55em}{}
\titlespacing*{\section}{0pt}{1.2em}{0.45em}
\titlespacing*{\subsection}{0pt}{1.0em}{0.35em}

\begin{document}

{\fontsize{16}{19}\selectfont\bfseries Understanding compression-extension error asymmetry in universal machine-learning interatomic potentials\par}
\vspace{0.45em}

{\fontsize{11}{14}\selectfont
Hongwei Du\textsuperscript{1,2},
Dingyang Lv\textsuperscript{1,2},
Boyan Zhang\textsuperscript{8},
Yu Ren\textsuperscript{3},
Feng Yu\textsuperscript{1,2},
Xin He\textsuperscript{1,2},
Bonan Zhu\textsuperscript{6},
Yongda Huang\textsuperscript{1,2},
Jiahui Liu\textsuperscript{1,2},
Yongheng Li\textsuperscript{1,2},
Siqi Shi\textsuperscript{5},
Jianjun Liu\textsuperscript{7},
Ziheng Lu\textsuperscript{1,2,3},
Hong Wang\textsuperscript{4,*}\par}

\vspace{0.35em}

{\fontsize{9}{11}\selectfont
1 Zhongguancun Academy, Beijing 100094, China\par
2 Zhongguancun Institute of Artificial Intelligence, Beijing 100094, China\par
3 Kairos Materials, Beijing, 100094, China\par
4 School of Materials Science and Engineering, Shanghai Jiao Tong University, Shanghai 200240, China\par
5 State Key Laboratory of Materials for Advanced Nuclear Energy \& School of Materials Science and Engineering, Shanghai University, Shanghai 200444, China\par
6 School of Aerospace Engineering, Beijing Institute of Technology, Beijing 100081, China\par
7 Shanghai Institute of Ceramics, Chinese Academy of Sciences, Shanghai 201899, China\par
8 School of Advanced Interdisciplinary Sciences, University of Chinese Academy of Sciences, Beijing, China\par
* Correspondence: Hong Wang (hongwang2@sjtu.edu.cn).\par
}
\vspace{0.55em}
\begin{quote}\small
\noindent Benchmarking of multicomponent materials links prediction errors in universal machine-learning interatomic potentials to local chemistry, geometric heterogeneity and interaction responses, with short contacts recurring in high-error environments. To understand how this short-distance error excess develops during deformation, we follow paired compression and extension paths from the same parent, comparing physical responses, prediction errors, geometry and model representations. We evaluate eleven pretrained potentials on 1,963 Materials Project trajectory (MPtrj) parents and 1,286 experimental-source MinX-Eq parents at six paired strain magnitudes from 1\% to 10\%. Density-functional theory (DFT) provides references for isotropic, uniaxial and biaxial loading in MPtrj and isotropic loading in MinX-Eq. Every model has a larger median force error under compression at \ensuremath{\pm}10\%, with pooled isotropic compression-to-extension ratios of 2.12 and 3.30 in the two cohorts. Relating error growth to DFT force-scale growth and relative-error evolution reveals different relative-accuracy trends beneath this common absolute-error pattern. From 1\% to 10\%, relative errors decrease on both branches, faster under compression in MPtrj and slower under compression in MinX-Eq. Compared with MPtrj, MinX-Eq has a weaker asymmetry in force-scale growth but a stronger asymmetry in error growth. Within-parent force-scale matching brings the error ratios toward unity, while MinX-Eq retains ratios of 1.108 to 1.472 across the models. Radial and local-environment analyses place compressed structures at shorter separations and farther from several public reference pools. Readout-associated model features show larger median displacements under compression in both cohorts. This shared structural and representation directionality accompanies different relative-error evolution, showing that greater environmental and feature displacement does not entail a uniform loss of relative accuracy. These results connect compression-extension error asymmetry to physical response amplitude, relative predictive accuracy, and structural and model responses, providing a basis for interpreting larger compression errors and assessing universal potentials in deformation simulations.
\end{quote}

\section{Introduction}
Universal machine-learning interatomic potentials learn energy and force relationships from large first-principles datasets, extending atomistic simulations across chemical compositions\cite{Deng2023CHGNet,BarrosLuque2026OMat24}. During structural relaxation, molecular dynamics and mechanical loading, local environments change continuously, and model reliability depends on describing the corresponding physical responses. Alongside aggregate predictive accuracy, identifying the environments in which errors arise and following their development as structures change are therefore important for understanding model applicability. Benchmarking of multicomponent materials has examined errors in relation to local chemistry, geometric heterogeneity and interaction-response characteristics and found that short contacts are a recurring feature of high-error environments\cite{Du2026Origins}. In addition, work on systematic softening has found that several universal potentials systematically underestimate energy and force responses away from low-energy configurations\cite{Deng2025Softening}. The former provides a structural lead on high errors at short contacts, while the latter identifies a possible systematic bias in model responses. Following errors as a structure is progressively compressed or extended connects these observations and examines how the short-distance error excess develops.
Paired compression and extension from the same parent allow the two branch errors to be followed from a common initial geometry with fixed atomic identities and composition. The first question is whether the error contrast persists across models and loading conditions, and how it develops with strain. Understanding its origin further requires distinguishing the strength of the physical response from changes in predictive accuracy. Equal-magnitude compression and extension can produce different reference forces, so a larger absolute error must be interpreted alongside the target force amplitude and examined for a corresponding difference in relative accuracy. How local environments and internal model representations evolve along the two paths, and how those changes relate to the errors, supply the structural and model-response context for interpreting the directional contrast.
Here we evaluate eleven pretrained potentials using 1,963 Materials Project trajectory (MPtrj) parents\cite{Deng2023MPtrj,Jain2013MaterialsProject} and 1,286 experimental-source MinX-Eq parents released with UniFFBench\cite{Mannan2025MinXEQ,Mannan2026UniFFBench}. The two collections provide different material settings for testing the same compression-extension error question. We apply six paired strains from 1\% to 10\% under isotropic loading to both collections, with additional uniaxial and biaxial paths for MPtrj. Density-functional theory provides reference energies, forces and stresses at the same geometries. We first establish the directional error pattern across strains, models and loading modes alongside the reference response. We then decompose error growth into force-scale growth and relative-error evolution, and compare errors at matched force scales to distinguish the roles of target amplitude and predictive accuracy. To relate prediction errors to environmental changes during deformation, we follow interatomic distances, departures of local environments from public reference pools, and displacements of readout-associated model features from the undeformed state along the same paths. These analyses examine whether larger compression errors accompany greater environmental departure and feature displacement, and whether these structural and representation asymmetries correspond to the same directional differences in relative predictive accuracy. By distinguishing stronger physical responses from changes in relative predictive accuracy, this study provides a basis for understanding compression-extension error asymmetry and assessing the predictive reliability of universal potentials along different deformation directions.
\section{Results}
\begin{figure}[!b]
  \centering
  \manuscriptfigure{0.52\textheight}{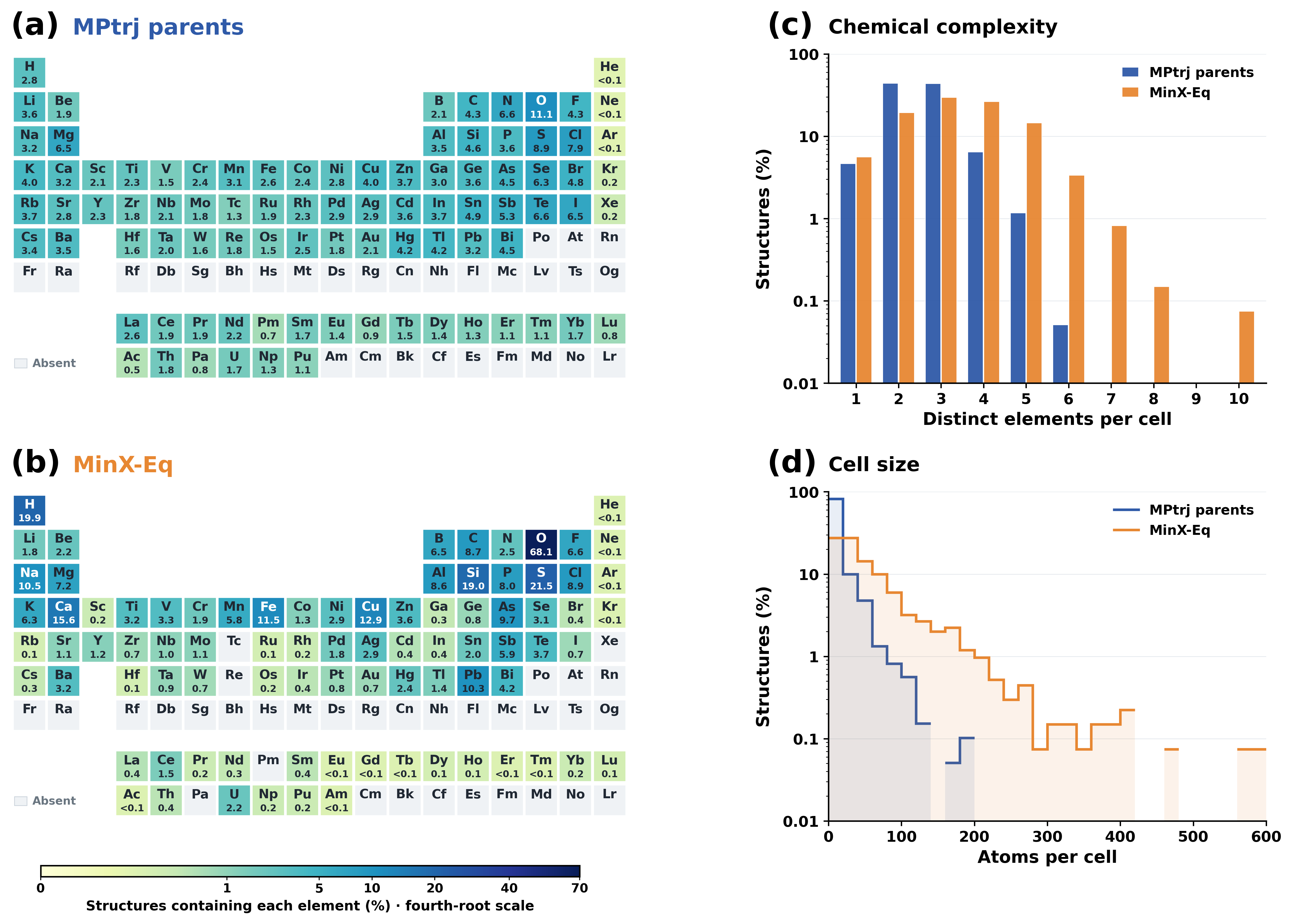}
  \caption{\textbf{Two complementary structure collections. }Element occurrence in (a) the 1,963 matched MPtrj parents and (b) all 1,344 experimental-source MinX-Eq records. Colors and tile labels give the percentage of source cells containing each element. Gray tiles indicate absent elements. Both periodic tables use a common fourth-root color normalization from 0\% to 70\%, with colorbar ticks labeled in actual percentages. (c) Number of distinct species per source cell. (d) Number of atoms per source cell in bins of 20 atoms. Both distributions show percentages on logarithmic vertical axes. Cells retain their original representation. The complete source collection in b differs from the completed DFT-model and representation cohorts defined in Methods.}
  \label{fig:1}
\end{figure}

The two structure collections provide different material settings for the paired comparison (Fig. 1). MPtrj covers a broad elemental range, with most cells containing two or three species and relatively few atoms. MinX-Eq contains more oxygen-bearing materials, more multicomponent cells and a longer tail of large cells. Oxygen occurs in 68.1\% of the MinX-Eq source structures and 11.1\% of the matched MPtrj parents. Compression and extension preserve each parent's composition and atomic identities, allowing the branch errors to be compared within the same material. The two collections test how this directional contrast varies across chemical and structural settings. DFT responses along the paired paths provide the target amplitudes against which the error differences can be assessed.

\begin{figure}[!htb]
  \centering
  \manuscriptfigure[0.87\linewidth]{0.74\textheight}{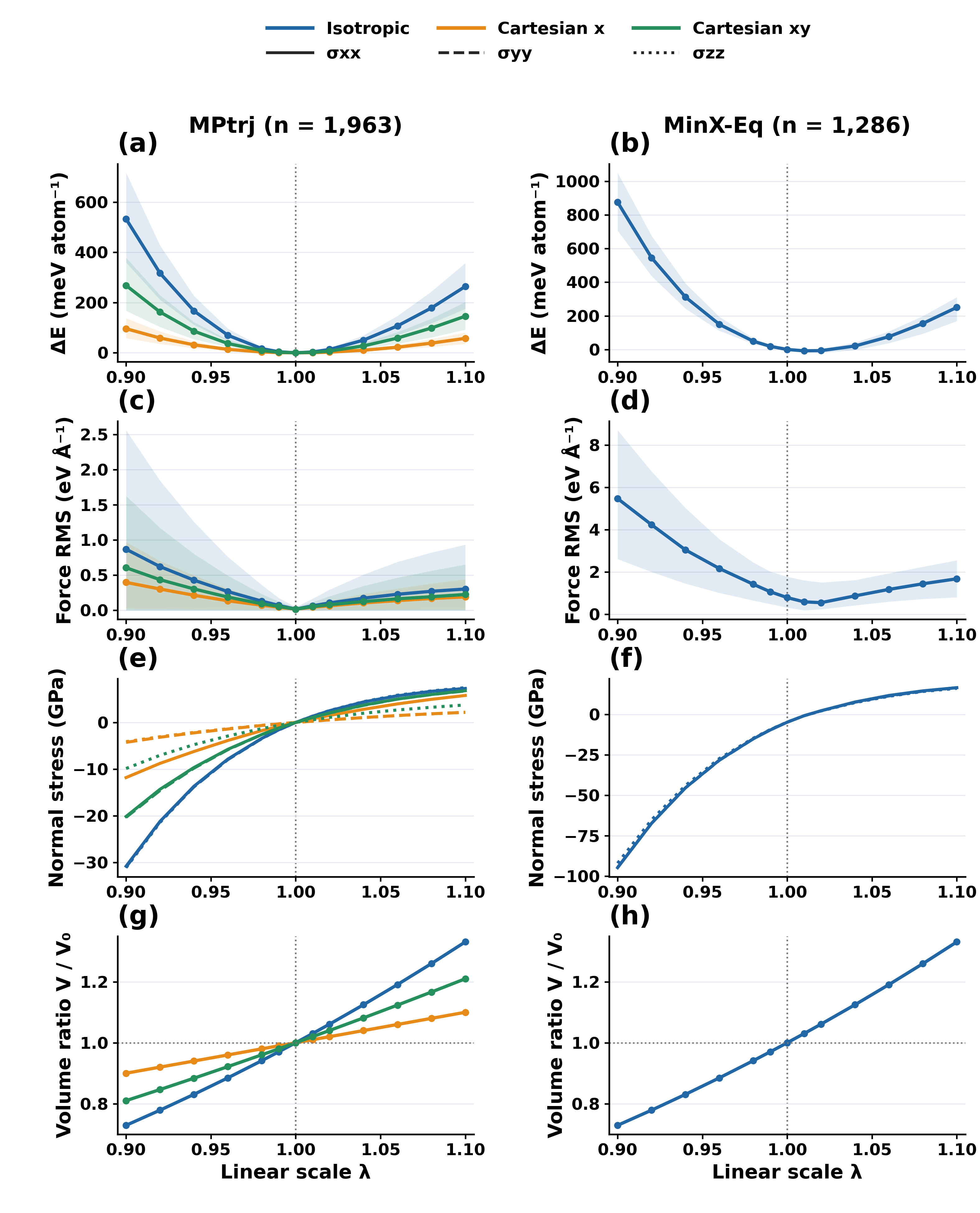}
  \caption{\textbf{Matched-parent DFT response in MPtrj and MinX-Eq. }The left column (a,c,e,g) shows 1,963 MPtrj parents under isotropic, Cartesian-x and Cartesian-xy loading. The right column (b,d,f,h) shows 1,286 MinX-Eq parents under isotropic loading. Rows show energy change per atom from the undeformed reference, RMS atomic force magnitude, tensile-positive normal stress components and V/V\textsubscript{0}. Curves are parent medians, and the energy and force bands are interquartile ranges. Fractional coordinates remain fixed. All panels use linear scale \ensuremath{\lambda} on the horizontal axis. Volume panels share vertical limits, while the other vertical scales are labeled for each cohort. Volume ratios are \ensuremath{\lambda}\textsuperscript{3}, \ensuremath{\lambda} and \ensuremath{\lambda}\textsuperscript{2} for isotropic, x and xy loading. Each path contains the actual \ensuremath{\lambda} = 1 geometry and twelve deformed states. The reference geometries can retain DFT forces and stress, as seen particularly in MinX-Eq.}
  \label{fig:2}
\end{figure}

Equal-magnitude compression and extension generate unequal physical responses in both cohorts (Fig. 2). At 10\% compression, the MPtrj median energy increases are approximately 533, 268 and 96 meV atom\textsuperscript{-}\textsuperscript{1} under isotropic, biaxial and uniaxial loading, with mode-dependent force and stress responses. For isotropic MinX-Eq, the compressed and extended endpoint energy changes are 876 and 250 meV atom\textsuperscript{-}\textsuperscript{1}, and the atomic force RMS values are 5.47 and 1.67 eV \AA{}\textsuperscript{-}\textsuperscript{1}. At the undeformed reference, MinX-Eq has a median force RMS of 0.785 eV \AA{}\textsuperscript{-}\textsuperscript{1}, and a small extension lowers its median energy. Its experimental-source geometries therefore begin away from DFT energy minima along the sampled path, giving the two cohorts different initial mechanical states. At \ensuremath{\pm}10\% isotropic strain, compression decreases the volume by 27.1\% and extension increases it by 33.1\%, so the stronger compression response accompanies the smaller volume excursion. Uniaxial loading gives equal absolute volume changes and allows a further comparison of directionality at equal volume excursions. Equal strain therefore leaves force and stress amplitudes direction dependent, and absolute prediction errors need to be interpreted alongside these target responses.
The model errors follow the direction of this physical asymmetry across all three MPtrj loading modes (Fig. 3). Every model has a larger median force error under compression at \ensuremath{\pm}10\%, with different overall accuracies across models. The pooled compression-to-extension force-error ratio increases from 1.39 under uniaxial loading to 1.73 under biaxial loading and 2.12 under isotropic loading. Stress-error ratios follow the same ordering, at 1.61, 2.34 and 3.18. Deforming more Cartesian directions accompanies a stronger error contrast, consistent with the ordering of the DFT response in Fig. 2. The contrast under uniaxial loading shows that error asymmetry also occurs at equal absolute volume excursions. Its presence in both atomic forces and stress places the larger compression error in predictions of both atomic and cell responses.

\begin{figure}[!htb]
  \centering
  \manuscriptfigure{0.78\textheight}{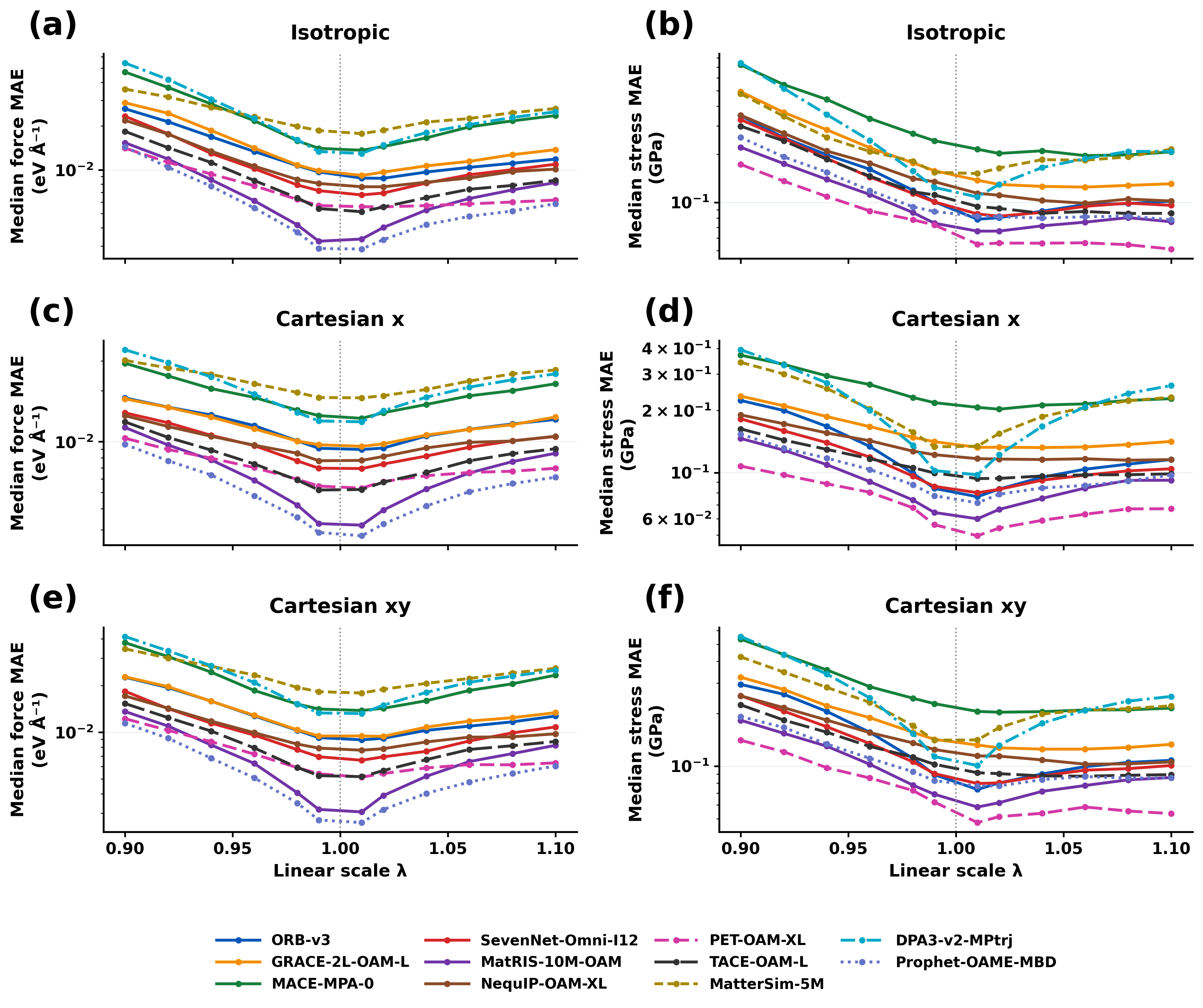}
  \caption{\textbf{Model errors across three MPtrj loading modes. }Median structure-level force-component MAE and six-component stress MAE for eleven pretrained models on the same 1,963 MPtrj parents. Rows correspond to isotropic, Cartesian-x and Cartesian-xy loading, with twelve deformation states per path. Horizontal axes show linear scale \ensuremath{\lambda}, and the line at \ensuremath{\lambda} = 1 marks the undeformed reference. Force and stress errors are plotted on logarithmic axes in eV \AA{}\textsuperscript{-}\textsuperscript{1} and GPa, respectively. Supplementary Figs. S2 and S3 give the matched-branch ratios.}
  \label{fig:3}
\end{figure}

Experimental-source structures show the same compression-side error excess, with a larger endpoint contrast (Fig. 4). At \ensuremath{\pm}10\% isotropic strain, the model-specific ratios of compression and extension median errors are 2.29 to 4.11 for forces and 3.15 to 8.70 for stresses. The pooled ratios reach 3.30 and 4.39, compared with 2.12 and 3.18 for isotropic MPtrj. Every model retains the compression-above-extension ordering of its median errors across the two populations with different compositions, cell sizes and initial mechanical states. Larger absolute compression errors are therefore shared by both cohorts, while the strength of the contrast varies with the material setting. Figure 2 also shows a stronger target response under compression. Comparing error growth with DFT force-scale growth tests whether the branch errors develop in proportion to the target amplitude and how changes in relative accuracy affect that contrast.

\begin{figure}[!htb]
  \centering
  \manuscriptfigure{0.78\textheight}{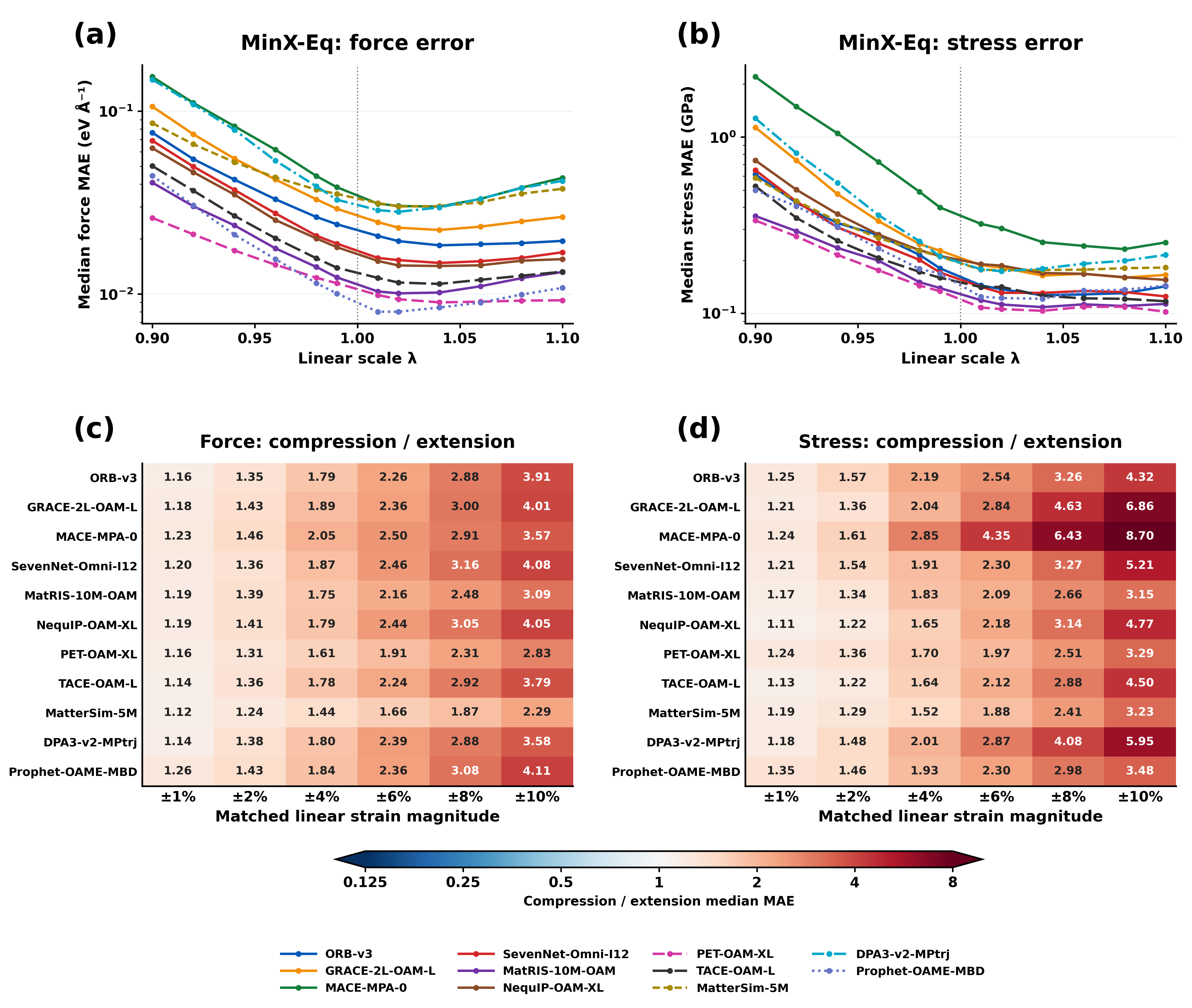}
  \caption{\textbf{Model errors and directional asymmetry in MinX-Eq. }(a,b) Median structure-level force and stress MAE for 1,286 MinX-Eq parents under isotropic loading. Horizontal axes show \ensuremath{\lambda}, and vertical axes show force MAE in eV \AA{}\textsuperscript{-}\textsuperscript{1} and stress MAE in GPa. (c,d) Compression median MAE divided by extension median MAE on the same parents at each absolute linear strain. Rows identify models and columns identify strains. Values above one indicate larger compression errors. The color scale is logarithmic from 1/8 to 8, with saturated values indicated and all ratios printed in the cells.}
  \label{fig:4}
\end{figure}

Force-scale decomposition expresses error growth as the product of DFT force-scale growth and relative-error change (Fig. 5). We define the DFT scale A as the mean absolute Cartesian force component and the relative error as q = e/A, where e is force MAE. On the fixed DFT-defined subsets, increasing strain from 1\% to 10\% raises the compression and extension force scales by 12.42 and 4.94 times in MPtrj and by 4.28 and 2.19 times in MinX-Eq (Fig. 5a,b). MPtrj has a compression-to-extension force-scale growth ratio of 2.514, error-growth ratios of 1.388 to 2.216 across models, and relative-error growth ratios of 0.552 to 0.881 (Fig. 5c). At \ensuremath{\pm}10\%, the geometric-mean compression-to-extension relative-error ratios are 0.527 to 0.854 in the same subset. The absolute compression errors are larger, but they occupy a smaller fraction of the target force scale. MinX-Eq has a force-scale growth ratio of 1.950, smaller than MPtrj, yet its error-growth ratios reach 2.038 to 3.197, with relative-error growth ratios of 1.046 to 1.640 (Fig. 5d). In both populations, geometric-mean relative errors decrease on both branches from 1\% to 10\%. The faster compression-side decrease in MPtrj makes error growth less asymmetric than force-scale growth. The slower compression-side decrease in MinX-Eq makes error growth more asymmetric than force-scale growth. Larger absolute compression errors can thus coexist with improving relative accuracy on both branches. The difference between the rates of improvement determines whether the force-scale growth asymmetry is attenuated or amplified in error growth.

\begin{figure}[p]
  \centering
  \manuscriptfigure[0.84\linewidth]{0.40\textheight}{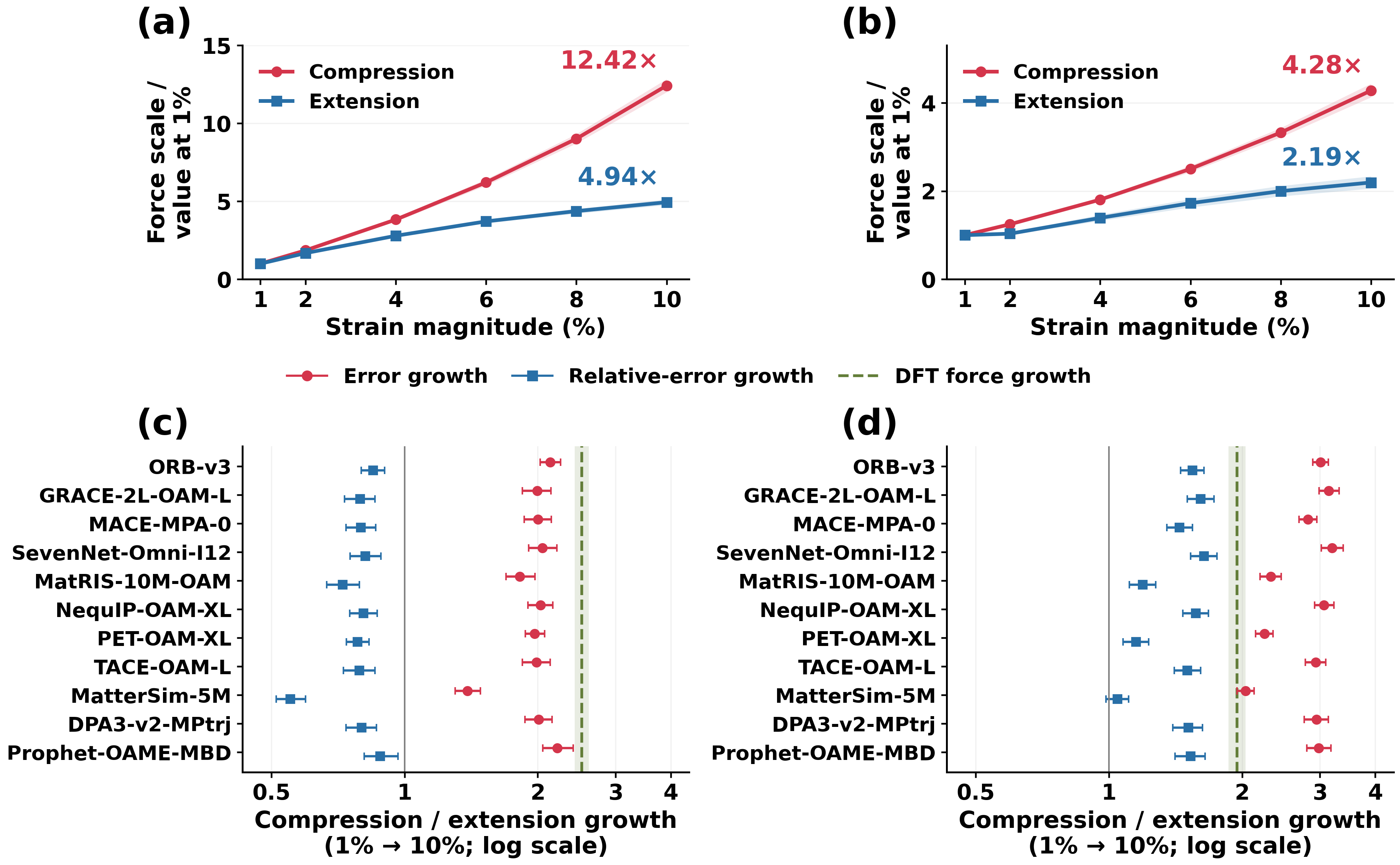}
  \caption{\textbf{Force-scale growth and error-growth decomposition. }Left and right columns show isotropic MPtrj and MinX-Eq results. (a,b) DFT force-component scale A = (3N)\textsuperscript{-}\textsuperscript{1} \textit{\ensuremath{\Sigma}}\textsubscript{i} \textit{\ensuremath{\Sigma}}\textsubscript{\ensuremath{\alpha}} |\textit{F}\textsubscript{DFT,i\ensuremath{\alpha}}| divided by its value on the same branch at 1\% strain. (c,d) Ratios of compression and extension growth factors from 1\% to 10\%. For X \ensuremath{\in} \{e,A,q\}, \textit{R}\textsubscript{X} = [\textit{X}\textsubscript{C}(10\%)/\textit{X}\textsubscript{C}(1\%)]/[\textit{X}\textsubscript{E}(10\%)/\textit{X}\textsubscript{E}(1\%)], where e is force MAE and q = e/A is relative error. Red, blue and green show \textit{R}\textsubscript{e}, \textit{R}\textsubscript{q} and \textit{R}\textsubscript{A}, respectively, with \textit{R}\textsubscript{e} = \textit{R}\textsubscript{A} \textit{R}\textsubscript{q}. All panels require A > 0.001 eV \AA{}\textsuperscript{-}\textsuperscript{1} at every strained state, selecting 1,364 of 1,963 MPtrj parents and 1,167 of 1,286 MinX-Eq parents. Estimates are parent-equal geometric means. Shading and error bars are pointwise 95\% intervals from 2,000 paired parent bootstrap samples. Supplementary Fig. S6 gives the within-parent force-scale matching comparison.}
  \label{fig:5}
\end{figure}

Matching the DFT force scale within each parent brings the branch errors closer together (Supplementary Fig. S6). Across the eleven models, compression-to-extension error ratios decrease to 0.986 to 1.202 in MPtrj and 1.108 to 1.472 in MinX-Eq. These geometric means use 1,108 and 391 parents, respectively, and compare the matched result with the equal-\ensuremath{\pm}10\%-strain result on the same parents. Matching generally pairs different strains on the two branches, changing the comparison from equal deformation amplitude to equal target force amplitude. Both cohorts move toward unity, while every MinX-Eq model retains a larger compression error after matching. Together, the growth decomposition and matched comparison show that the equal-strain error contrast contains directional differences in both target amplitude and relative accuracy.
The force scale describes the amplitude of the prediction target, while radial distributions describe the interatomic geometry that produces this response (Fig. 6). At small strain, both branches lie near the central region of the distance distributions in three public reference pools. Increasing strain moves compression toward shorter distances and extension toward longer distances in both cohorts. The pools differ in distribution width and shape and consequently in radial coverage of the deformed structures. Each deformation path thus changes both the target response amplitude and the distance region sampled by the model inputs. Radial distributions place the larger compression forces and errors in regions of shorter interatomic contacts. Local environments also change through neighboring species and their spatial arrangement, whose departure from public reference structures is further quantified by SOAP distances.

\begin{figure}[p]
  \centering
  \manuscriptfigure[0.84\linewidth]{0.40\textheight}{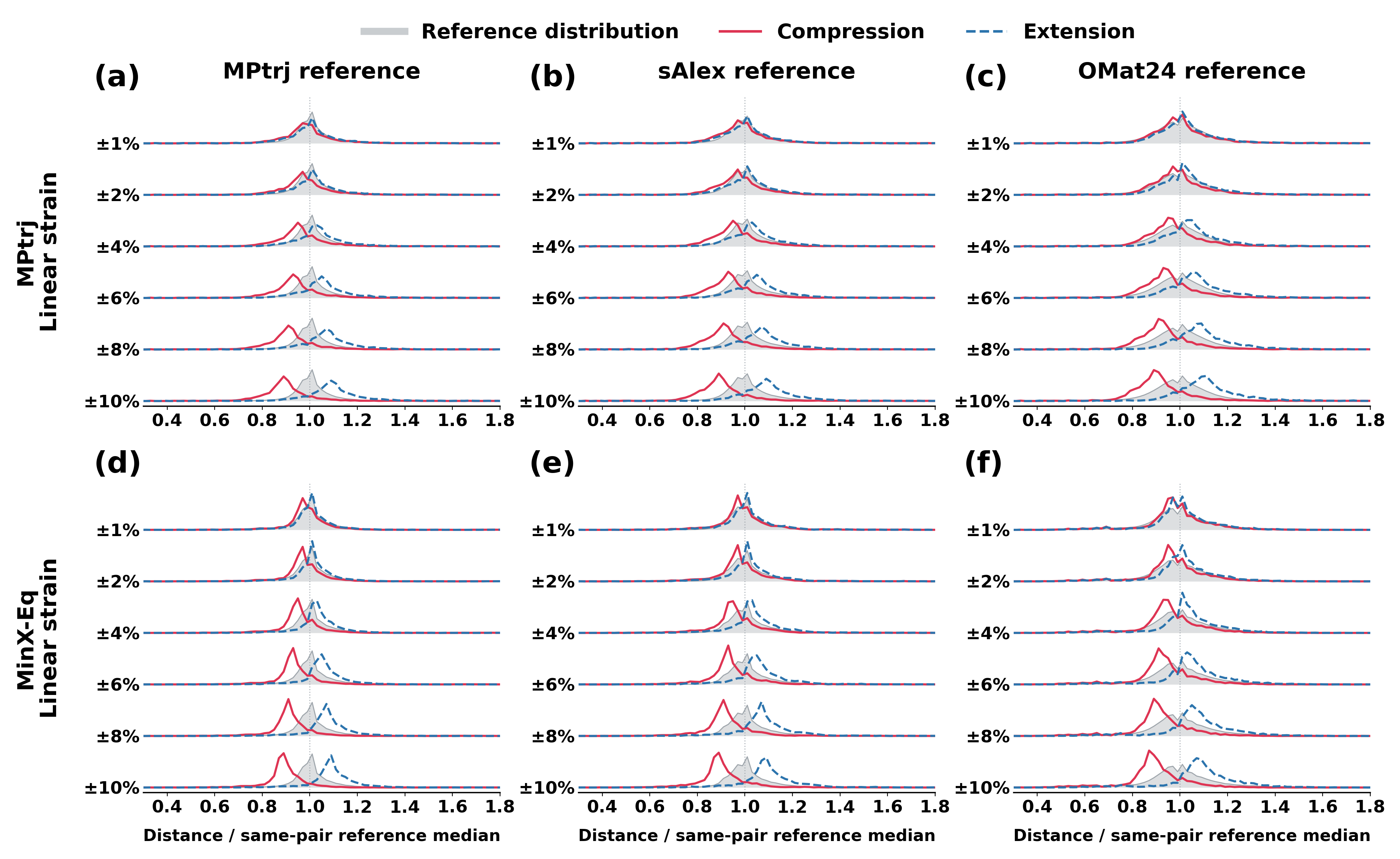}
  \caption{\textbf{Radial environments relative to three public references. }Radial distributions for MPtrj (upper row) and MinX-Eq (lower row) at six paired strain magnitudes, with MPtrj, sAlex and OMat24 references in the three columns. The horizontal coordinate is pair distance divided by the same-element-pair reference median. Vertically offset curves show probability densities at the labeled strains. Gray curves show the reference, red solid curves show compression, and blue dashed curves show extension. Each reference uses a common density-height multiplier across strains and cohorts. Densities are normalized over the full distance range.}
  \label{fig:6}
\end{figure}

SOAP distances place compressed environments farther from public reference environments than extended environments in both cohorts (Fig. 7)\cite{Bartok2013SOAP,Himanen2020DScribe,Laakso2023DScribe}. Against MPtrj references, the median structural distances for compression, near-reference and extension states are 0.091, 0.0087 and 0.030 in MPtrj and 0.145, 0.0226 and 0.0528 in MinX-Eq. Compression exceeds extension for 93.1\% and 96.3\% of paired parents. The median endpoint ordering persists against sAlex, with compression/extension distances of 0.0776/0.0203 and 0.1493/0.0686, and against OMat24, with 0.0385/0.0070 and 0.0893/0.0331. OMat24 supplies closer reference environments on both branches while preserving their directional ordering. Changing the reference pool changes a structure's reference distance while leaving its prediction error fixed. Comparable distances within each state span broad error ranges, and different models give different errors for structures with identical reference distances. Reference distance therefore measures geometric proximity to the selected public environments, with a model-dependent relation to prediction error. The larger environmental departure under compression is shared by both populations, while their relative-error evolution in Fig. 5 has different branch orderings.

\FloatBarrier
\begin{figure}[H]
  \centering
  \includegraphics[width=1.0\linewidth,height=0.85\textheight,keepaspectratio]{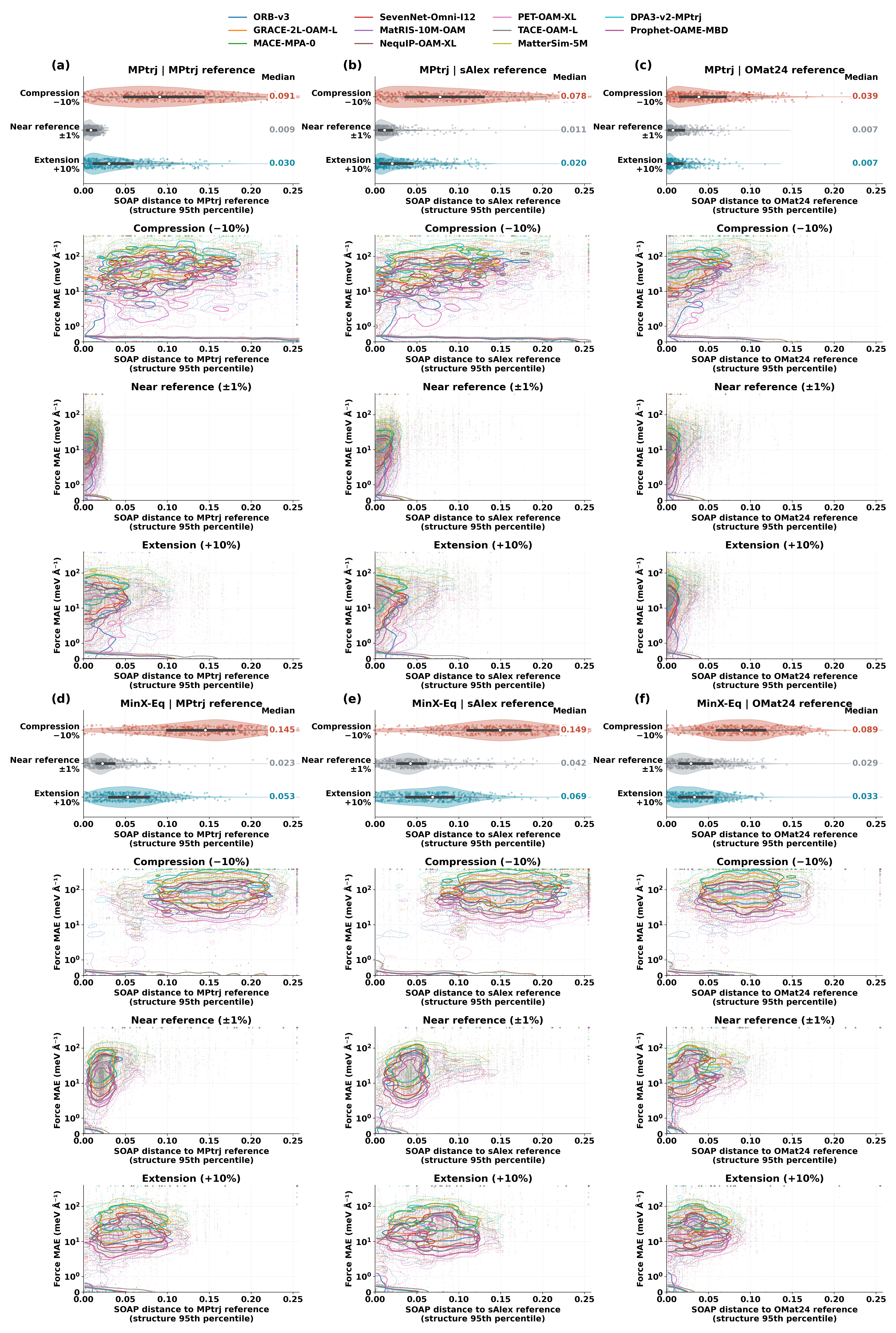}
  \caption{\textbf{Local-reference distances and force errors. }Six groups compare 1,963 MPtrj parents (a,b,c) and 1,286 MinX-Eq parents (d,e,f) against MPtrj, sAlex and OMat24 SOAP reference pools from left to right. Each labeled group contains four vertically stacked plots: structural-distance distributions for \ensuremath{-}10\%, pooled \ensuremath{\pm}1\% and +10\%, followed by force-error comparisons for compression, near-reference and extension states in that order. Structural distance is the 95th percentile of atomwise nearest-neighbor distances to references with the same central element. The near-reference group pools the \ensuremath{\pm}1\% states. Figure 8 uses the actual \ensuremath{\lambda} = 1 reference. All observations enter the statistics and densities, and boundary markers locate observations beyond the common plotting limits. The joint panels plot structural SOAP distance against force MAE in meV \AA{}\textsuperscript{-}\textsuperscript{1}, with contours at the 50\% and 80\% density levels.}
  \label{fig:7}
\end{figure}

\begin{figure}[!b]
  \centering
  \includegraphics[width=1.0\linewidth,height=0.75\textheight,keepaspectratio]{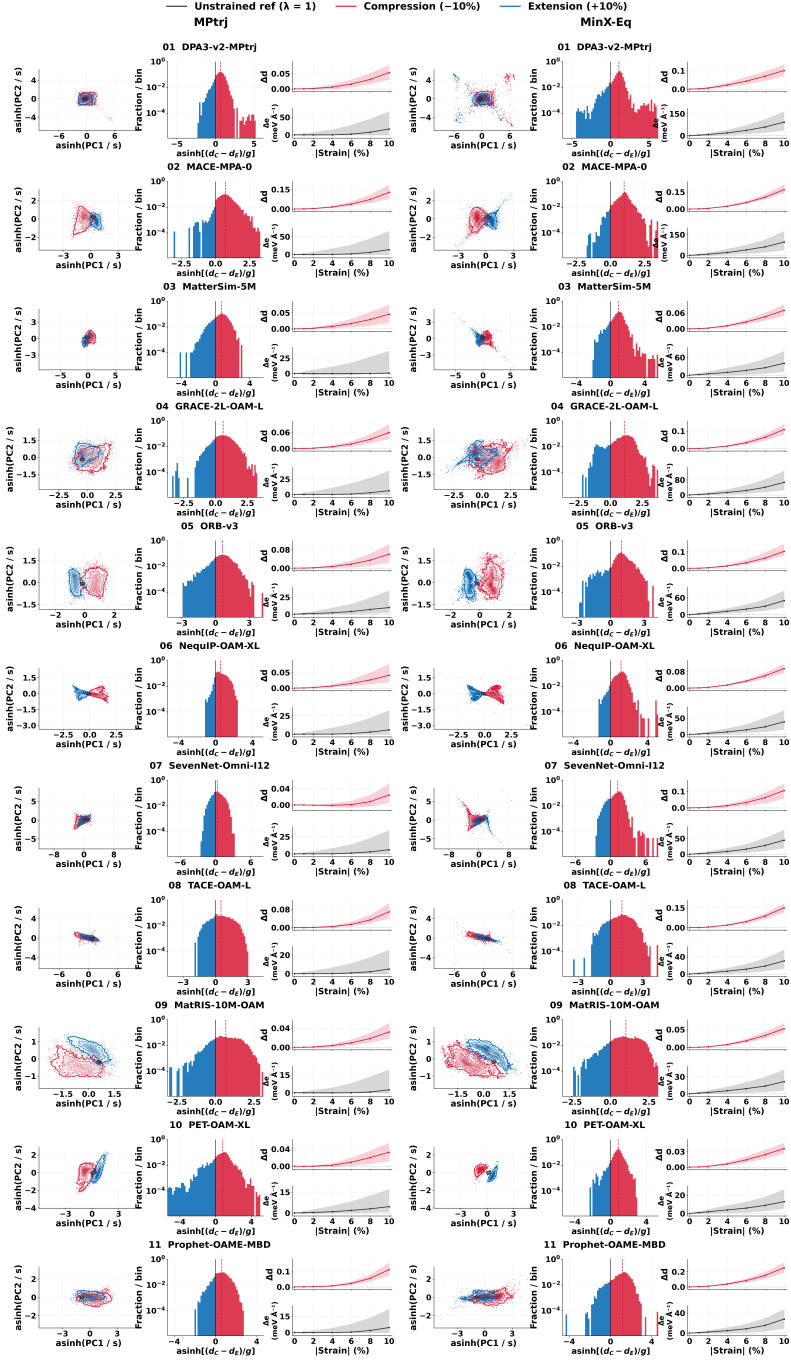}
  \caption{\textbf{Representation and force-error asymmetry along matched strain paths. }Features are extracted at model-specific stages associated with the readout. Each model uses 1,963 MPtrj and 1,285 MinX-Eq parents after the fixed exclusions. Left: response-centered joint PCA of the actual reference, \ensuremath{-}10\% and +10\% atomic features, shown in gray, red and blue. A common within-model asinh scale is used, contours enclose 50\%, 80\% and 95\% of probability mass, and points retain the tails. PC1 and PC2 explain 10.3\% to 87.6\% of feature variance across models. Middle: distributions of asinh[(\textit{d}\textsubscript{C} \ensuremath{-} \textit{d}\textsubscript{E})/g], where \textit{d}\textsubscript{s} = \ensuremath{\Vert}\textit{h}\textsubscript{s} \ensuremath{-} \textit{h}\textsubscript{ref}\ensuremath{\Vert}\textsubscript{2}/S uses all extracted feature dimensions, S is the undeformed feature RMS and g is the pooled weighted median absolute gap. The logarithmic vertical axis gives the parent-weighted atomic fraction per bin. Red and blue bins show positive and negative gaps, and both cohorts share g within each model. Right: median paired representation gap \ensuremath{\Delta}d and force-error gap \ensuremath{\Delta}e in meV \AA{}\textsuperscript{-}\textsuperscript{1} against absolute linear strain. Bands span the 25th to 75th percentiles among parents. Differences are computed within parents before aggregation. Each panel has its own labeled linear vertical scale. Dotted, dashed and solid PCA contours enclose 50\%, 80\% and 95\% of probability mass, respectively. In the histograms, the black solid line marks zero and the red dashed line marks the weighted median. Positive gaps indicate larger compression responses. Model names identify the rows.}
  \label{fig:8}
\end{figure}

Readout-associated atomic features also show a larger compression-side response within the models (Fig. 8). Relative to each structure's undeformed state, compression produces larger full-dimensional displacements for 65.2\% to 96.4\% of parent-weighted atomic environments in MPtrj and 85.8\% to 97.9\% in MinX-Eq. All twenty-two model and cohort combinations have a positive median distance difference, showing that the common geometric directionality also appears in features used for readout. The representation gap generally increases with the paired force-error gap along the strain paths. For ORB-v3 from 1\% to 10\%, the median representation gap increases from 0.00040 to 0.0605 in MPtrj and from 0.00098 to 0.103 in MinX-Eq, while the force-error gaps rise from 0.44 to 9.69 and from 2.78 to 49.95 meV \AA{}\textsuperscript{-}\textsuperscript{1}. Near the initial geometry, the onset of the two gaps varies across models. SevenNet has a slightly negative representation gap in MPtrj, and some models develop a representation gap before appreciable growth of the median force-error gap. Environmental departure, feature displacement and absolute error therefore share a compression-side directionality at large strain, with differences in their onset. In conjunction with Fig. 5, larger feature displacements accompany faster relative-accuracy improvement under compression in MPtrj and slower improvement under compression in MinX-Eq. Feature displacement and relative predictive accuracy are thus distinct response characteristics. Their comparison reveals common directionality in feature displacement alongside material-dependent evolution of relative accuracy.

\FloatBarrier
\section{Discussion}
Paired deformation of the same parent, with chemical identity fixed, establishes larger absolute compression errors across eleven pretrained interatomic potentials. This trend persists across three MPtrj loading modes and in experimental-source MinX-Eq structures, including uniaxial loading with equal absolute volume excursions. The larger isotropic and biaxial error contrasts accompany stronger DFT responses, so the directionality of errors needs to be understood together with that of the target response. Force-scale decomposition shows that larger absolute compression errors can coexist with improving relative accuracy. In MPtrj, the compression branch has smaller endpoint relative errors and a faster relative-error decrease with strain, making error growth less asymmetric than force-scale growth. Relative errors also decrease on both MinX-Eq branches, but the slower compression-side decrease makes error growth more asymmetric than force-scale growth. Matching the force scale brings the branch errors closer in both cohorts, while MinX-Eq retains compression-to-extension ratios of 1.108 to 1.472. Absolute errors measure the force deviations encountered in simulation, and relative errors measure those deviations against the target amplitude. Together, they characterize directional predictive reliability. Structural and representation comparisons connect this interpretation to the deformation paths. Compression shortens contacts, moves local environments farther from several public reference pools in SOAP space, and produces larger median displacements of readout-associated features. From input geometry to model representations, this common directionality accompanies different relative-accuracy trends, showing that greater environmental or feature displacement does not entail a uniform loss of relative accuracy. The spread of errors at comparable reference distances and the different onset of feature and error gaps in some models further distinguish the evolution of environmental change, internal response and force-prediction error. Understanding compression-extension error asymmetry therefore requires tracking the amplitude of the growing physical response, the error relative to that response, and the accompanying structural and representation changes. Their joint evolution along paired paths distinguishes absolute-error growth alongside stronger physical responses from directional differences in relative predictive accuracy, providing a basis for assessing universal potentials in deformation simulations. Future work should develop interpretable models of the compression-extension error gap to distinguish its structural and model-related origins and guide improvements to the environmental representations of interatomic potentials for more accurate predictions along both deformation directions.
\section{Methods}
\subsection{Structures and deformation protocol}
MPtrj parents come from the Materials Project trajectory dataset\cite{Deng2023MPtrj}, and MinX-Eq comes from the MinX-EQ archive in Zenodo record 16733258\cite{Mannan2025MinXEQ}. The composition comparison uses 1,963 matched MPtrj parents and all 1,344 MinX-Eq source records, counted in their original cells. The MPtrj calculations began with 2,000 parents, of which 1,963 have complete data for the three loading modes. In MinX-Eq, 1,342 records were prepared for calculation after excluding eq\_0131 and eq\_0683. The intersection with complete DFT and model error pairs for all eleven models and twelve isotropic deformation states contains 1,286 parents. For a cell matrix H with lattice vectors as rows, the deformed cell is H\ensuremath{^{\prime}} = H D and atomic positions are r\ensuremath{^{\prime}} = sH\ensuremath{^{\prime}} at fixed fractional coordinates s. The matrices D are diag(\ensuremath{\lambda},\ensuremath{\lambda},\ensuremath{\lambda}), diag(\ensuremath{\lambda},1,1) and diag(\ensuremath{\lambda},\ensuremath{\lambda},1) for isotropic, Cartesian-x and Cartesian-xy deformation. The last two scale Cartesian components of the cell, including in non-orthogonal cells. The definitions were checked against HTEM\cite{Yang2025HTEM}. The twelve scales are 0.90, 0.92, 0.94, 0.96, 0.98, 0.99, 1.01, 1.02, 1.04, 1.06, 1.08 and 1.10. The actual undeformed geometry at \ensuremath{\lambda} = 1 supplies the DFT and representation reference. The representation analysis applies the fixed exclusions eq\_0369, eq\_0139, eq\_0156, eq\_1200 and eq\_1213 to every model and strain. The last four are already absent from the complete force-label cohort, so removing eq\_0369 leaves 1,285 MinX-Eq parents for Fig. 8. Representation distributions, force labels and paired trends use the same structure identities. Together, the two representation cohorts contain 89,047 atoms, with all atoms in the retained structures, including zero feature vectors, entering the analysis.
\subsection{DFT calculations and protocol consistency}
Static DFT calculations use the Materials Project-compatible VASP\cite{Kresse1996VASP} workflow implemented in quacc\cite{Rosen2026QuaccSoftware,Rosen2026QuantumAccelerator} and historical PAW-PBE potentials\cite{Blochl1994PAW,Kresse1999PAW,Perdew1996PBE} identified by stored hashes. MPtrj calculations with source-task records inherit their material-specific electronic inputs, including documented solver translations between VASP versions. These results take priority when available. The remaining historical calculations retain their recorded settings, including a fixed EDIFF of 10\textsuperscript{-}\textsuperscript{7} eV, with provenance records distinguishing the calculation protocols. MinX-Eq uses the quacc MP legacy static setup derived from MPRelaxSet\cite{Ong2013Pymatgen}, with material-specific input manifests and historical PAW-PBE potentials. Its electronic tolerance is derived from EDIFF\_PER\_ATOM = 5 \ensuremath{\times} 10\textsuperscript{-}\textsuperscript{5} eV. All 16,718 DFT points for the 1,286 parents and thirteen states are checked for unchanged atom counts, finite labels and the intended isotropic cell scaling. Reference and deformed inputs are compared within each parent to confirm identical EDIFF and POTCAR settings. An independent set of 160 structures tests reproduction of the original MPtrj labels using exact historical PAW-PBE potentials (Supplementary Fig. S5). Mean absolute differences in energy, force and stress are 0.001059 eV atom\textsuperscript{-}\textsuperscript{1}, 0.01153 eV \AA{}\textsuperscript{-}\textsuperscript{1} and 0.2191 GPa. The corresponding medians are 7.204 \ensuremath{\times} 10\textsuperscript{-}\textsuperscript{5} eV atom\textsuperscript{-}\textsuperscript{1}, 0.001063 eV \AA{}\textsuperscript{-}\textsuperscript{1} and 0.009626 GPa, quantifying protocol agreement for the reproduced structures. The reproducibility archive includes input manifests, solver records and source hashes.
\subsection{Pretrained models and error definitions}
We evaluate the pretrained ORB-v3\cite{Rhodes2025ORBv3}, GRACE-2L-OAM-L\cite{Lysogorskiy2026GRACE}, MACE-MPA-0\cite{Batatia2025MACE,MACE2026FoundationModels}, SevenNet-Omni-I12\cite{Kim2026SevenNet}, MatRIS-10M-OAM\cite{Zhou2026MatRIS}, NequIP-OAM-XL\cite{Kavanagh2026Equivariant}, PET-OAM-XL\cite{Bigi2026PET}, TACE-OAM-L\cite{Xu2026TACE}, MatterSim-5M\cite{Yang2024MatterSim}, DPA3-v2-MPtrj\cite{Zhang2026DPA3} and Prophet-OAME-MBD\cite{Kairos2026Prophet} releases with their weights fixed. Supplementary Table S2 lists checkpoint identities, release identifiers and asset hashes. For parent p, model m and state s, force MAE is \textit{e}\textsubscript{F}(p,m,s) = (3\textit{N}\textsubscript{p})\textsuperscript{-}\textsuperscript{1} \textit{\ensuremath{\Sigma}}\textsubscript{i} \textit{\ensuremath{\Sigma}}\textsubscript{\ensuremath{\alpha}} |\textit{F}\textsubscript{ML,i\ensuremath{\alpha}} \ensuremath{-} \textit{F}\textsubscript{DFT,i\ensuremath{\alpha}}|. Figures 3 and 4 report force MAE in eV \AA{}\textsuperscript{-}\textsuperscript{1}. Figure 7 reports force MAE in meV \AA{}\textsuperscript{-}\textsuperscript{1}, which is also the unit for paired force-error differences in Fig. 8 and the text. Stress MAE averages absolute differences over the six saved Voigt components in GPa, with tensile-positive signs after conversion. The DFT force-response measure in Fig. 2 is the atomic RMS magnitude [\textit{N}\textsubscript{p}\textsuperscript{-}\textsuperscript{1}\textit{\ensuremath{\Sigma}}\textsubscript{i}\ensuremath{\Vert}\textit{F}\textsubscript{DFT,i}\ensuremath{\Vert}\textsubscript{2}\textsuperscript{2}]\textsuperscript{1}/\textsuperscript{2} in eV \AA{}\textsuperscript{-}\textsuperscript{1}. Energy response is the DFT energy change from the undeformed reference divided by the atom count. The paired force-error difference is \ensuremath{\Delta}e(p,m,\ensuremath{\delta}) = \textit{e}\textsubscript{F}(p,m,1\ensuremath{-}\ensuremath{\delta}) \ensuremath{-} \textit{e}\textsubscript{F}(p,m,1+\ensuremath{\delta}), computed within each parent before aggregation. Ratio plots use either mean(\textit{e}\textsubscript{C})/mean(\textit{e}\textsubscript{E}) or median(\textit{e}\textsubscript{C})/median(\textit{e}\textsubscript{E}), as specified in their captions. Pooled ratios divide the two medians after combining structure-model observations on each branch. Interquartile bands describe variation among parents. At zero strain the two states coincide, giving a paired gap of zero by definition.
\subsection{Force-scale accounting and matched comparison}
For Fig. 5, the DFT force scale is A(p,s) = (3\textit{N}\textsubscript{p})\textsuperscript{-}\textsuperscript{1} \textit{\ensuremath{\Sigma}}\textsubscript{i} \textit{\ensuremath{\Sigma}}\textsubscript{\ensuremath{\alpha}} |\textit{F}\textsubscript{DFT,i\ensuremath{\alpha}}|, and relative error is q(p,m,s) = e(p,m,s)/A(p,s). A averages absolute net-force components over atoms and Cartesian directions using the same convention as force MAE, while Fig. 2 uses atomic force RMS. Figures 3 and 4 compare absolute errors across the full cohorts. Figure 5 requires A > 0.001 eV \AA{}\textsuperscript{-}\textsuperscript{1} at all twelve strained states, selecting 1,364 MPtrj and 1,167 MinX-Eq parents for every branch, strain and model comparison. For X \ensuremath{\in} \{e,A,q\}, the growth ratio is \textit{R}\textsubscript{X} = [\textit{X}\textsubscript{C}(10\%)/\textit{X}\textsubscript{C}(1\%)]/[\textit{X}\textsubscript{E}(10\%)/\textit{X}\textsubscript{E}(1\%)]. The identity \textit{R}\textsubscript{e} = \textit{R}\textsubscript{A} \textit{R}\textsubscript{q} holds for each parent and for the parent-equal geometric means exp[mean(log \textit{R}\textsubscript{X})]. Sensitivity analyses use thresholds of 0.0001, 0.01 and 0.1 eV \AA{}\textsuperscript{-}\textsuperscript{1}. Force-scale matching requires \textit{A}\textsubscript{C} and \textit{A}\textsubscript{E} to increase strictly over the six strain magnitudes and to have overlapping ranges. These conditions retain 1,108 MPtrj and 391 MinX-Eq parents. Each parent's target is the midpoint of its common log A interval. Piecewise linear interpolation of log e against log A within the two observed ranges gives the matched error ratio, generally at different strains on the two branches. The equal-strain comparison uses the same parents at \ensuremath{\pm}10\%. Both ratios are summarized by parent-equal geometric means. Targets at 25\% and 75\% of the common log A interval provide sensitivity checks. Shading and error bars in Fig. 5 and Supplementary Fig. S6 are pointwise percentile 95\% intervals from 2,000 paired bootstrap samples of parents, using seed 20261008. All eligible errors, including the tails, contribute with their original values.
\subsection{Public-reference environment comparisons}
Radial distances are divided by the corresponding same-element-pair reference median. Reference and target densities retain the original analysis weights and use one display-height multiplier per reference, shared across cohorts and strains. The displayed window is 0.3 to 1.8, with densities normalized over the full radial range. Periodic SOAP descriptors\cite{Bartok2013SOAP,Himanen2020DScribe,Laakso2023DScribe} use unit normalization, a 4 \AA{} cutoff, n\_max = 2, l\_max = 2, Gaussian width 0.4 \AA{} and mu1nu1\cite{Darby2022Compression} compression on a common 118-element basis. Nearest-neighbor searches compare atoms with the same central element. The public reference pools comprise the full MPtrj cache, 93 sAlex files and 5,300 OMat24 sublibraries. sAlex is a curated subset of Alexandria\cite{Schmidt2023PhaseDiagram,MetaFAIR2026OMat24Data}. Pool construction follows public-dataset membership independently of the evaluated checkpoints, so the distances measure environmental proximity to these public collections. Ranking uses float32 with TF32 disabled, and distances to the selected vectors are recomputed in float64. Near-neighbor identities can be numerically tied at very small distances. Structural distance is the 95th percentile of atomwise nearest-reference distances. The near-reference group pools atoms from the \ensuremath{\pm}1\% branches before taking this percentile and averages their force errors. All parents contribute to the density estimates and statistics. Common plotting limits are used in Fig. 7, with boundary markers for observations outside the displayed window.
\subsection{Model representations and paired distances}
We extract atomic features at model-specific stages associated with the readout. ORB, MACE, MatterSim and Prophet contribute 256 dimensions each. SevenNet, MatRIS and DPA3 contribute 128, NequIP and TACE contribute 320, GRACE contributes 34, and PET contributes 3,200. Distances use all coordinates of these selected features. Atom identities, feature shapes and source hashes are checked before assembly, and ORB uses the same extraction version for reference and deformed states. Model-specific feature definitions and forward-consistency checks accompany the data. The PCA response vector is \textit{x}\textsubscript{s,i} = [\textit{h}\textsubscript{s,i} \ensuremath{-} (\textit{h}\textsubscript{0.99,i} + \textit{h}\textsubscript{1.01,i})/2]/S, with the actual \textit{h}\textsubscript{1.00,i} transformed by the same expression. The RMS scale S comes from undeformed feature norms with equal cohort and parent weighting and preserves variation in feature magnitude among atoms. One PCA is fitted per model to the full covariance, with equal weights for the two cohorts and three displayed states, equal parent weights within each cohort, and equal atom weights within each parent. The concentration of the reference distribution partly reflects centering on the paired \ensuremath{\pm}1\% feature midpoint. Within each model, both PCA coordinates share an asinh display scale, and all states and cohorts share axis bounds and density-estimation rules. Contours enclose 50\%, 80\% and 95\% of the distribution in the transformed display coordinates. A common Gaussian bandwidth is used, with grid refinement to keep enclosed-mass discretization errors below 0.5 percentage points. All retained atoms are displayed, with points outside the 95\% contour emphasized. PCA projections display the geometry of the feature response, and displacements from the undeformed state are calculated in the full extracted feature space.
The full-dimensional distance is \textit{d}\textsubscript{s,i} = \ensuremath{\Vert}\textit{h}\textsubscript{s,i} \ensuremath{-} \textit{h}\textsubscript{1.00,i}\ensuremath{\Vert}\textsubscript{2}/S, measured from the actual undeformed state. Histograms show \textit{d}\textsubscript{C,i} \ensuremath{-} \textit{d}\textsubscript{E,i} at \ensuremath{\pm}10\% with parent-equal atomic weights. Both cohorts share the model-specific asinh scale g, defined as the pooled weighted median absolute gap. All samples and tails contribute. Red bins indicate positive differences and blue bins indicate negative differences. For strain-path summaries, \textit{d}\textsubscript{s,p} is the mean atomic distance within parent p. We form \textit{\ensuremath{\Delta}d}\textsubscript{p} = \textit{d}\textsubscript{C,p} \ensuremath{-} \textit{d}\textsubscript{E,p} before taking the cohort median and 25th to 75th percentiles at each magnitude. Each model and cohort has its own labeled linear vertical scale. The numerical archive contains complete atomwise distances, parent summaries and extreme values. Independent verification replays every force label, checks histogram probability masses and 44 trend summaries, and recomputes the twelve deformed-state distances for 55 model-parent cases directly from raw features.
\section*{Data availability}
\addcontentsline{toc}{section}{Data availability}
MPtrj and MinX source data are available from the cited repositories\cite{Deng2023MPtrj,Mannan2025MinXEQ}. The public reference collections are described in Refs.\cite{BarrosLuque2026OMat24,MetaFAIR2026OMat24Data}. The study archive contains parent identifiers, calculation provenance, force and stress labels, reference distances and representation paths. Public deposition of these study-specific data is pending.
\section*{Code availability}
\addcontentsline{toc}{section}{Code availability}
The study archive contains the code for deformation generation, model registration, feature extraction, reference searches and figure generation, together with source versions and figure checksums. Public deposition of this code is pending.
\FloatBarrier
\addcontentsline{toc}{section}{References}
\IfFileExists{reference.bib}{%
  \bibliographystyle{unsrtnat}
  \bibliography{reference}
}{%
  \begin{quote}\small\itshape
  The external bibliography database \texttt{reference.bib} is unavailable in this isolated preview.
  \end{quote}
}

\end{document}